\documentclass[aps,pra,10pt,showpacs,twocolumn,floatfix,superscriptaddress,nofootinbib,hyperref]{revtex4-1}
\usepackage{amsmath}
\usepackage{amssymb}
\usepackage[T1]{fontenc}
\usepackage{graphicx}
\usepackage{longtable}

\newcommand{\dd}{\mathrm{d}}
\newcommand{\ee}{\mathrm{e}}
\newcommand{\ii}{\mathrm{i}}

\begin{document}

\title{Quantum annealing speedup of embedded problems via suppression of Griffiths singularities}

\author{Sergey Knysh}
\affiliation{
USRA Research Institute for Advanced Computer Science (RIACS)}
\author{Eugeniu Plamadeala}
\affiliation{
USRA Research Institute for Advanced Computer Science (RIACS)}
\affiliation{Quantum Artificial Intelligence Laboratory (QuAIL), NASA Ames Research Center
}
\author{Davide Venturelli}
 \email{davide.venturelli@nasa.gov}
\affiliation{
USRA Research Institute for Advanced Computer Science (RIACS)}
\affiliation{Quantum Artificial Intelligence Laboratory (QuAIL), NASA Ames Research Center
}

\begin{abstract}
Optimal parameter setting for applications problems embedded into
hardware graphs is key to practical quantum annealers (QA). Embedding
chains typically crop up as harmful Griffiths phases, but can be used as
a resource as we show here: to balance out singularities in the logical
problem changing its universality class. Smart choice of embedding
parameters reduces annealing times for random Ising chain from
$O(\exp[c\sqrt N])$ to $O(N^2)$. Dramatic reduction in time-to-solution
for QA is confirmed by numerics, for which we developed a custom
integrator to overcome convergence issues.
\end{abstract}

\maketitle

Implementation of quantum 
annealing~\cite{kadowaki1998quantum,farhi2000quantum,*farhi2001quantum,das2008colloquium,albash2018adiabatic}  
for optimization problems expressed as
unconstrained quadratic forms of binary variables requires 
embedding~\cite{choi2008minor,*choi2011minor,hilton2019systems}
the connectivity topology of the problem into that of the underlying hardware.
This typically means representing logical qubits in terms of clusters of
physical qubits designed to be ferromagnetically aligned at the completion of
the annealing cycle. Performance suffers as a result: Not only does it lead to a
decreased utilization of qubits, but since the effective two-level system for
strongly coupled qubits is endowed with slower dynamics, it becomes more
susceptible to harmful non-adiabatic excitations. Judicious choice of embedding
parameters to minimize these effects has large practical utility and remains
an active area of research.
Most of the earlier work studied optimal parameter setting
empirically by running experiments on D-Wave annealers
as well as formulating educated guesses based on classical and quantum spectrum
and intuition from the theory of phase transitions~\cite{king2014algorithm,venturelli2015quantum,fang2019minimizing,harris2018phase}.

The adiabatic quantum computing protocol 
obtains a solution by evolving, in a finite time
interval $[0;T]$, a time-dependent Hamiltonian that interpolates between a
``driver'' term with easily obtainable ground state and the ``classical''
Hamiltonian which is diagonal in Z-basis and encodes the cost of underlying
quadratic unconstrained binary optimization (QUBO) problem:
\begin{equation}
  H(t) = -\bigl(1-\tfrac{t}{T}\bigr) \sum_i X_i - \tfrac{t}{T}
    \Biggl(\sum_{\langle i k \rangle} J_{ik}Z_i Z_k+\sum_i h_i Z_i \Biggr).
  \label{qubo}
\end{equation}
The time-dependent Hamiltonian describes the annealing of Ising spins
on a graph as the external uniform magnetic field $\Gamma =
\frac{1-s}{s}$ (where $s = \frac{t}{T}$) is applied in the transverse
direction and decreases to zero.
Adiabatic theorem ensures that for sufficiently large $T$ the system
remains close to its instantaneous ground state until the driver term
that causes bit flips vanishes at the end of the annealing algorithm. Very recently it has been shown theoretically that even a simple driver such as the transverse field could in principle achieve a superpolynomial speedup against classical computing~\cite{hastings2020power}.

Known empirical evidence suggests that high probability of success is best
achieved by performing many independent runs over optimally chosen annealing
cycle time $T$. The probability of error $\epsilon=(1-p_0)^L$ can be made
arbitrarily small using large number of repeats $L$ (here $p_0$ is the success
probability of a single run). The total computation time $\mathcal{T}= TL$ is
expressed in terms of unambiguous time-to-solution metric
$\mathcal{T}= \tau \log (1/\epsilon)$ that we adopt here:
\begin{equation}
  \tau = \min_T \frac{T}{\bigl| \log [1-p_0 (T)] \bigr|} .
  \label{tts}
\end{equation}
Although we restrict ourselves to a linear schedule with driver term and
classical term prefactors $A(s)=1-s$ and $B(s)=s$ for the sake of
simplicity, singularities at $s=0$ and $s=1$ dictate the scaling of $p_0(T)$
only for very large $T$. Using a sweet-spot value of $T$ ensures that our
results remain qualitatively robust for a general non-linear interpolating
schedule typically constrained by a specific hardware implementation.

The illustrative example that is the subject this letter is a simple model:
a one-dimensional chain of ferromagnetically coupled Ising spins (qubits) with
random interactions,
\begin{equation}
  H(s)=(1-s)H_D+sH_C, \text{ with } H_C=-\sum_{i = 1}^N J_i Z_{i-1}Z_i
  \label{qa}
\end{equation}
(in zero longitudinal field, $h_i=0$) with the usual choice of a driver term 
that represents a transverse field, $H_D=-\sum_i X_i$.
The annealing bottleneck for this problem is related to a
quantum critical point separating paramagnetic and ferromagnetic phases.
Without randomness, i.e. if all $J_i=1$, the energy gap that separates the
ground state from excited states within the symmetric subspace is minimized at
$s_c=1/2$ ($\Gamma_c=1$). Notice that we disregard one half of the
states (including the first excited state) that are never populated due to $H(t)$
preserving a global symmetry $\left[ U=\exp \frac{\pi\ii}{2}\sum_i X_i\right]$
which flips all bits: $Z_i \rightarrow -Z_i$. The critical exponents
that can be obtained either from an analytic solution via fermionization or
from the renormalization group (RG) analysis predict that correlation length
$\xi \sim |s-s_c|^{- 1}$ and the gap in the paramagnetic phase
$\Delta E \sim (s_c-s)$. This suggests a cutoff value of the gap
$\Delta E_c \sim 1/N$ as the correlation length $\xi$ reaches the
system size $N$, and, therefore, polynomial scaling for TTS metric
$\tau \sim (\Delta E_c \cdot \Delta s_c)^{-1} \sim N^2$. This scaling is
intimately related to Kibble-Zureck mechanism of quench dynamics
across the phase transition~\cite{zurek2005dynamics,dziarmaga2005dynamics}. 

By contrast, if couplings $J_i$ are randomly distributed and are not
correlated, the RG flow is toward an infinite-randomness fixed point (IRFP)
with the gap $\Delta E \sim \exp(-c\sqrt{\xi})$, i.e. the minimum gap
scales with $N$ as a stretched exponential~\cite{fisher1992random,*fisher1995critical}.
This anomalous scaling is due to the presence of Griffiths-McCoy
singularities~\cite{griffiths1969nonanalytic,*mccoy1969incompleteness}
as different parts of the system are unable to reach criticality
simultaneously as a result of strong disorder fluctuations.

To illustrate, consider uniform distribution of the couplings
$J_k \in [0;1]$. The critical value of the transverse field is the
geometric mean, i.e. the gap to the 2$^\text{nd}$ excited state
is minimized at $\Gamma_c=1/\ee$
(corresponding to $s_c=(1+1/\ee)^{-1} \approx 0.731$). If the chain
is cut in two equal parts at criticality, the subchains will have
$\Gamma_c^{1,2}=\Gamma_c \pm \Delta \Gamma_c$
where $\Delta \Gamma_c \sim 1/\sqrt{N}$ by central limit theorem.
One part will be ferromagnetically ordered with gap
$\Delta E_c \sim \exp{(-c N \Delta \Gamma_c)}$. This only demonstrates
stretched exponential scaling for the gap to the \emph{first} excited state,
which is not relevant to the transverse field parity-conserving dynamics.
However, it is expected that entire low-energy spectrum has
the same universal scaling form (see Appendix~\ref{app:Eigenspectrum}
for a rigorous discussion). In practice, the asymptotic scaling
will not set in until $N$ becomes moderately large, whereas scaling
for small system sizes may be indistinguishable from that of a
non-disordered chain. 

\paragraph{Embedding using block spins.}

For this study we set out to explore strategies for setting parameters for
embeddings in coherent quantum annealers, which replace logical qubits
with blocks of $M$ ferromagnetically coupled physical qubits. This is
done primarily to increase the connectivity of qubit interaction graph
[to $M(c-2)+2$, where $c$ is the degree of connectivity graph of
physical hardware]. The extreme example achieves all-to-all
connectivity at the cost of quadratic reduction of the number of
logical qubits [to $O\bigl(\sqrt{N}\bigr)$] as seen in
Refs.~\onlinecite{choi2008minor, choi2011minor}.
Another important application is extending the range of Ising couplings,
which is more relevant for the present study of a linear chain ($c=2$).
Lastly, using block spins has been suggested as a form of
error-correction~\cite{pudenz2014error,*pudenz2015quantum,%
vinci2015quantum,matsuura2017quantum}.

\begin{figure}[!t]
  \includegraphics[width=\linewidth]{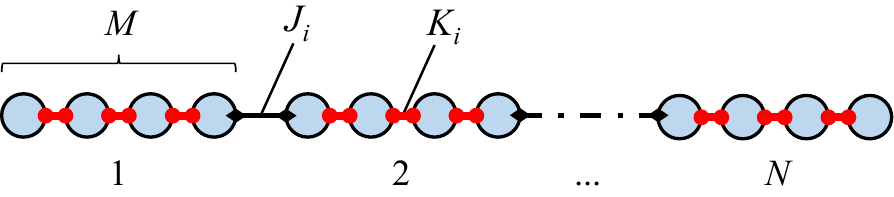}
  \caption{\label{fig:model}
    A simple model of embedding. A logical problem corresponds to
    1D spin chain with ferromagnetic couplings $J_i$. Embedded problem
    replaces each logical spin with a block (chain) of $M$ physical
    spins having intra-chain coupling $K_i$.
  }
\end{figure}

As we introduce $M$ ancillary spins for each logical variable
(see  Fig.~\ref{fig:model}), the classical Hamiltonian reads
\begin{gather}
  H_C = -\sum_{k=1}^{NM-1} \tilde{J}_k Z_{k-1} Z_k, \label{emb} \\
\text{with} \quad \tilde{J}_{Mi}=J_i, \quad 1 \leqslant i \leqslant N-1, \notag \\
\text{and} \quad \tilde{J}_{Mi+1}=\cdots=J_{Mi+(M-1)}=K_i,
\quad 0 \leqslant i \leqslant N-1. \notag
\end{gather}
For simplicity, the ferromagnetic couplings $K_i$ within a subchain are taken
to be uniform but we allow variations from one logical qubit to another. To
ensure that these ferromagnetic links are never broken in any \emph{local}
minimum we insist that
\begin{equation}
  \min_{0 \leqslant i \leqslant N - 1} K_i > \max_{1 \leqslant i \leqslant N - 1} J_i.
  \label{constr}
\end{equation}
Since the model retains its 1D character, existing analytical techniques
are still applicable and numerical studies can explore a regime where
annealing becomes intractable for large $N$.

Specifically, we observe that for sufficiently small values of the transverse
field $\tilde{\Gamma} < \min_i K_i$ we may perform the real-space
renormalization procedure due to D.~S.~Fisher~\cite{fisher1992random,*fisher1995critical}.
Each subchain now behaves as a spin subjected to a transverse field
with the renormalized value
\begin{equation}
  \Gamma_i = \frac{\tilde{\Gamma}^M}{\prod_{k=Mi+1}^{Mi+M-1}\tilde{J}_k}
  = \frac{\tilde{\Gamma}^M}{K_i^{M-1}} .
  \label{Gi}
\end{equation}
(In a reversal of notation used in Ref.~\onlinecite{fisher1992random,*fisher1995critical},
we write $\tilde{\Gamma} = \frac{s}{1-s}$ to represent the bare value
of the transverse field and reserve non-accented letters $\Gamma_i$ to
denote the renormalized fields.) The effective model is still an Ising chain
\begin{equation}
  H = s \left[ -\sum_{i=1}^{N-1} J_i Z_{i - 1} Z_i -
    \sum_{i=0}^{N-1} \Gamma_i X_i \right],
  \label{renorm}
\end{equation}
albeit with the local transverse field that can potentially have spatial
variations.

The embedding procedure thus unlocks an important resource since many existing
implementations either lack such local control entirely or have limited
dynamic range. It allows us to restore the polynomial scaling of
annealing complexity, balancing out disorder fluctuations to
synchronize the phase transition across entire chain.
We make the following ansatz for embedding parameters
\begin{equation}
  K_i = C (J_i J_{i+1})^{-\frac{1}{2(M-1)}} \quad (i = 1, \ldots, N-2)
  \label{Ki}
\end{equation}
for some rescaling constant $C.$ At the edges we can use, e.g., $K_0=C J_1^{-1/(M+1)}$
and $K_{N-1}=C J_{N-1}^{-1/(M+1)}$.

With this choice, renormalized local fields in the bulk become $\Gamma_i =
\frac{\tilde{\Gamma}^M}{C^{M-1}} \sqrt{J_i J_{i + 1}}$ so that fluctuations of the
value of the critical field $\Delta \tilde{\Gamma}_c$ for large subchains now
scale as $1/N$.

\begin{figure}[!th]
  \includegraphics[width=\linewidth]{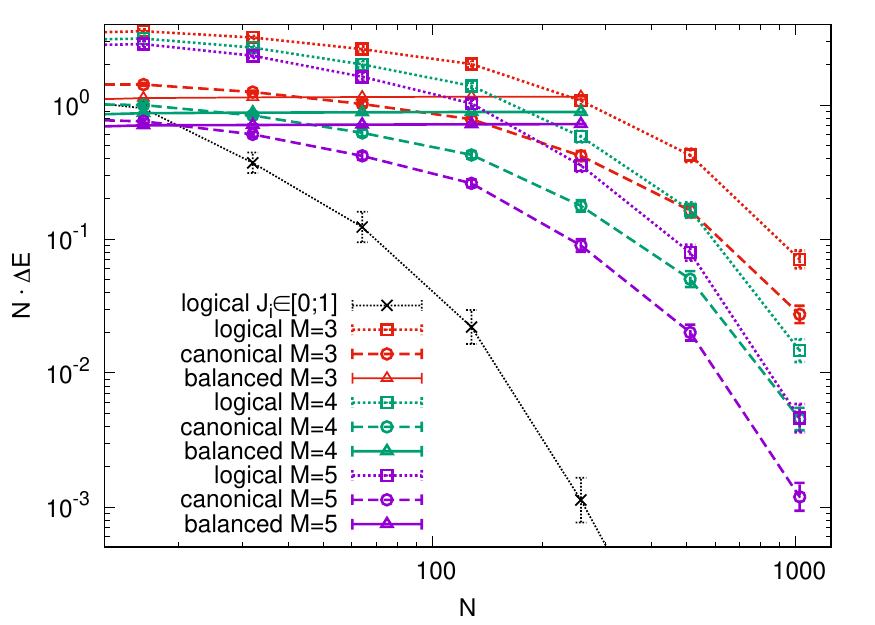}
  \caption{\label{fig:gap}
    Normalized median value of the minimum gap ($N \cdot \Delta E_c$) vs problem size. 
    Black solid line corresponds to the logical problem
    with strong disorder ($J_i \in [0;1]$). Results for embedded
    variants with $M=3,4,5$ are shown with solid and dashed color
    lines for, respectively, the balanced choice of embedding parameters
    given by scaled-down form of Eq.~(\ref{Ki}) and the canonical choice 
    $K_i=1$. The disorder distributions used for embedded variants are
    those uniform in the interval $[1/M;1]$. For reference, the median
    values of the minimum gap of the logical problem for these weaker
    disorder distributions are also shown using dotted color lines.
  }
\end{figure}

In Fig.~\ref{fig:gap} we compare the median minimum gap for different parameter setting prescriptions, showing that the balanced choice in Eq.~(\ref{Ki}) recovers a polynomial gap.
It is imperative to discuss how the parameters of our numerical study
were chosen. The black solid line plots the estimate of
the median value (taken over disorder distribution) of the minimum gap
for strong disorder: $J_i$ are taken to be uniformly distributed in
the interval [0;1]. The data clearly cannot be fit by any power law
(which would have been a straight line on a log-log plot).

It is evident from Eq.~(\ref{Ki}) that allowing arbitrarily small values
of $J_i$ would in turn lead to arbitrarily large values of $K_i$.
Moreover, small values of $J_\text{min}$ are not expected to appear
naturally for Ising representations of practical
problems~\cite{lucas2014ising} and would make the problem particularly
prone to misspecification errors in hardware annealers.

To mitigate this problem, we choose $J_i \in [1/M;1]$, where $M$ is the
number of spins in a block. Fig.~\ref{fig:gap} uses colored dotted line to
plot gaps of the logical problem for such disorder distributions.
Dashed lines show the gaps for an embedded problem with
a canonical choice of embedding  parameters ($K_i=1$).
Balanced choice of embedding parameters from Eq.~(\ref{Ki}) with $C=1$ would have
yielded values $K_i$ in a range $[1;M^{1/(M-1)}]$. Current practice
for existing hardware implementations of quantum annealing is to use
largest possible values of couplings $J_i$ in order to minimize
misspecification error, hence it is natural to assume
that $J=1$ already represents the maximum programmable value.
In order to accommodate the balanced parameter setting, all 
ferromagnetic couplings must be scaled down by a factor of 
$M^{1/(M-1)}$, i.e. $J_i \in [M^{-M/(M-1)};M^{-1/(M-1)}]$ and
$C=M^{-M/(M-1)^2}$.
This rescaling step had been made for the
data plotted in Fig.~\ref{fig:gap} using solid colored lines. The motivation
for this peculiar choice of disorder distribution is that it maximizes
the range $J_\text{max}-J_\text{min}$ of inter-chain couplings after the rescaling.

Rescaling of couplings parameters downward handicaps quantum annealing
performance. Using rescaled $J'=\lambda J$ we can show that
$\Delta E_c'\cdot\Delta s_c' = (\lambda^2 s_c/s'_c) \Delta E_c \cdot \Delta s_c$.
The scaling factor can be rewritten as $(1-s_c)\lambda^3+s_c\lambda^2$,
which is less than unity if $\lambda<1$. Numerical data for the gap is
consistent with this qualitative behavior. More rigorous analysis of
the low-energy spectrum along the lines described in
Appendix~\ref{app:Eigenspectrum} also predicts that the gap for
canonical embedding is smaller than that of the logical problem with
the same disorder by $O(M)$ factor, also in agreement with the numerical
data.

We encountered a problem specific to instances using balanced embedding,
likely due to a quirk in its spectrum: numerical diagonalization could
only performed for a smaller range of $N$ as the limits of machine
precision were reached. It may be possible to overcome this obstacle
using multiprecision arithmetic to find the roots of characteristic
polynomial for the tridiagonal matrix. Fortunately, the median gap fits
polynomial scaling perfectly (straight line on a log-log plot), which
gives us sufficient confidence to extrapolate the data for large $N$.

\paragraph{Time-to-solution analysis.}

The one-dimensional Ising chain is the simplest example of an integrable quantum
model. A rotation ($X_i \rightarrow -Z_i$, $Z_i \rightarrow X_i$) followed by
a Jordan-Wigner transformation expresses the Hamiltonian as a quadratic form
in fermionic operators. The solution of time-dependent Bogolyubov-de Gennes
(BdG) equations is used to write these operators in Heisenberg representation
(which is a linear combination of Schr\"odinger operators since $H$ is
quadratic~\cite{dziarmaga2006dynamics,caneva2007adiabatic}). For
instance, using Majorana representation (see e.g. Ref.~\onlinecite{fradkin2013field})
($\chi_{2i}=X_i\prod_{k=0}^{i-1} Z_k$ and $\chi_{2i+1}=Y_i\prod_{k=0}^{i-1}Z_k$)
we can write
\begin{equation}
  \boldsymbol\chi(t) =\mathbf{S}(t) \boldsymbol\chi (0), \quad
  \frac{\dd \mathbf{S}}{\dd t} =2 \mathbf{M} \mathbf{S}.
  \label{BdG}
\end{equation}
Here $\mathbf{S}$ is special orthogonal $2n \times 2n$ matrix, whereas
$\mathbf{M}$ is skew-symmetric and also happens to be tridiagonal with
the elements on the upper diagonal $M_{2i,2i+1}=1-s$ and
$M_{2i-1, 2i}=s J_{i-1}$. This special structure further aids numerics;
on top of logarithmic reduction in complexity afforded by the
free-fermion mapping.

Eigenvalues of $\mathbf{M}$ (e.g. for ordering $\lambda_{2k}=+\ii\epsilon_k$,
$\lambda_{2k+1}=-\ii\epsilon_k$ with $0<\epsilon_0<\cdots<\epsilon_{N-1}$)
completely determine the spectrum of the instantaneous Hamiltonian $H(t)$
as a sum of single-particle excitations:
$E_{\{n_k\}} = 2 \sum_k n_k \epsilon_k$ where $n_k \in \{0,1\}$.
Specifically, the gap between the ground state and the second excited state
(which is second lowest in energy among those in the symmetric subspace) is
$\Delta E=2(\epsilon_1+\epsilon_2)$. This value enters the adiabatic condition
used to estimate the runtime (see Appendix~\ref{app:Eigenspectrum}).

To better quantify the performance we concentrate on a TTS metric of Eq.~(\ref{tts}).
The probability to find the system in its ground state at the end of the
annealing cycle at $t = T$ is
\begin{equation}
  p_0 = \biggl\langle \prod_k (b_k b_k^\dag) \biggr\rangle,
  \label{p0}
\end{equation}
where $b_k$ and $b_k^\dag$ are fermionic quasiparticle operators in Dirac
representation that diagonalize the final classical Hamiltonian $H_{C}$.
Expectation value can be computed at time $t=0$ provided that
Heisenberg representation for $b_k(t)$ and $b_k^\dag(t)$ is used.
The initial state satisfies $a_i|0\rangle=0$ where $a_i$ and $a_i^\dag$
are the Dirac fermion operators that diagonalize the transverse field Hamiltonian $H_D$.

Since the Hamiltonian is quadratic, the expectation value (\ref{p0}) is computed by
performing the sum over all possible pairings and can be expressed as
\begin{equation}
  p_0 = \sqrt{\det \tfrac{1}{2}(\mathbf{S}\mathbf{A}\mathbf{S}^T +\mathbf{B})}
  \label{p0det}
\end{equation}
where the only non-zero elements of $\mathbf{A}$, $\mathbf{B}$ are
$A_{2k,2k+1}=B_{2k-1,2k}=B_{0,2N-1}=1$ and $A_{2k+1,2k}=B_{2k,2k-1}=
B_{2N-1,0}=-1$ (see Appendix~\ref{app:groundstate}).

\paragraph{Numerical results.}

Investigating the complexity using the TTS metric is somewhat
challenging numerically. It entails integrating systems of differential
equations of more than $10^4$ variables for multiple choices of
annealing times (in excess of $10^7$ in dimensionless units).
Numerical error accumulates proportionally to the evolution time,
and is further amplified when we evaluate the determinant (\ref{p0det}).
After multiple tests, we implemented an integrator based on
Cayley transform and Magnus expansion~\cite{blanes2000improved,*blanes2002high} 
up to 8th order.
However, we use Pad\'e approximation of the exponential
to obtain a straightforward mapping to Cayley transform and
exploit the tridiagonal structure of matrix $\mathbf{M}$ to attain
best performance. 
Compared with traditional Runge-Kutta approaches used for these
types of problems~\cite{mbeng2019dynamics}, our method maintains
orthogonality of $\mathbf{S}$ at all update steps which 
significantly improves stability for long integration times
(see Appendix~\ref{app:numericalproc}).
For a given timestep our implementation outperforms comparable 
methods such as Dormand-Prince algorithms that are the method of
choice in celestial mechanics~\cite{dormand1980family}, but also
maintains excellent precision for large step sizes $\Delta t$.

Note that since optimal annealing time in Eq.~(\ref{tts}) cannot
be known in advance, it seems unfair to perform optimization of 
for each individual instance. We use ensemble-wide $T_\text{opt}(N)$,
which is determined through minimization, as a function of annealing
time, of the \emph{median} TTS, from a sample of random instances of
the same size. This choice is similar to the practice of benchmarking
of application problems.

\begin{figure}[!ht]
  \includegraphics[width=\linewidth]{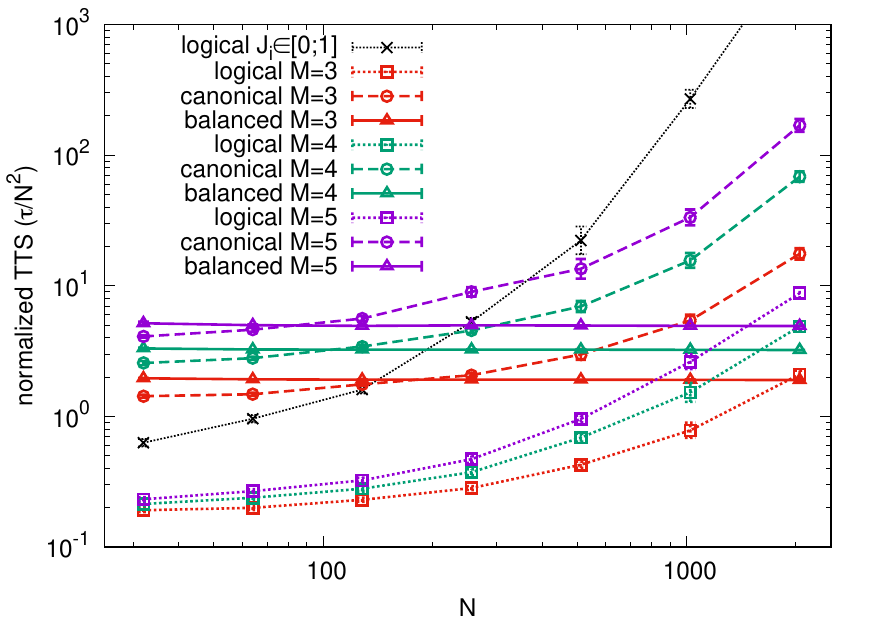}
  \caption{\label{fig:tts}
    Normalized median time-to-solution (divided by $N^2$) as a
    function of problem size. Block solid line plots TTS for a logical
    instance with strong disorder for $J\in[0;1]$. Dotted color lines
    represent logical instances for a disorder in the interval $[1/M;1]$
    for $M=3,4,5$. TTS for embedded variants using the same disorder
    distribution for logical couplings using the canonical and
    balanced choice of parameters are plotted using dashed and solid
    color lines, respectively.
    Errorbars represent the uncertainty of estimating the median
    using a small sample size of instances and optimization over 
    annealing time.
  }
\end{figure}

We plot the median time-to-solution for random instances of increasing
size in Fig.~\ref{fig:tts}. As before, colored dotted lines refer to
logical problem with $J_i \in [1/M;1]$, dashed lines correspond to
canonical embedding parameters and solid lines present results for
balanced embedding parameters. Finally, black solid line refers to a
logical problem with strong disorder $J_i \in [0;1]$, where
non-polynomial behavior is most pronounced. We observe that for 
the largest sizes balanced embedding clearly outperforms the canonical
one ($K_i=1$).

\paragraph{Conclusions}

We have investigated quantum annealing of embeddings of a simple
problem, where logical qubits were replaced by ferromagnetic chains
of Ising spins. A novel ansatz for a balanced choice of coupling
parameters based on renormalization group intuition results in an
exponential improvement in annealing times. This is corroborated
numerically, using time-to-solution metric of complexity.
For large sizes the balanced choice significantly outperforms both
canonical parameter setting and the quantum annealing of the logical
problem using the same distribution of disorder (but with no embedding).

The protocol proposed in this letter succeeds because embedding
parameters are set in such a fashion that spatially separate regions
achieve criticality simultaneously. A situation were multiple domains
oriented in random directions are created is thus avoided.
Remarkably, this can be accomplished without local control of the
transverse field. Complementary approaches described in literature
rely on ``growing'' the domain and require significant modifications
to quantum annealing protocol using time-dependent local control of
the transverse field~\cite{rams2016inhomogeneous}.

An important venue of research is the generalization of this idea to
higher-dimensional problems. Dearth of exactly solvable models
suggests using approximate methods to find embedding parameters
that achieve synchronization of local phase transitions. We expect to
be able to suppress Griffiths singularities relieving the phase
transition bottleneck of annealing complexity. However, Ising models
in higher dimensions exhibit frustration, which dominates complexity
for very large instances~\cite{knysh2016zero} (crafted problems in 1D
can also exhibit frustration bottleneck~\cite{roberts2020noise}).
We are cautiously optimistic that this general technique can be useful
for intermediate-range problems that could be implemented in highly coherent quantum annealers~\cite{novikov2018exploring}. Another research direction is the
extension of this work to open systems where the annealing dynamics
is incoherent, driven by coupling with an external reservoir~\cite{smelyanskiy2017quantum,mishra2018finite}. 
This would open up the opportunity of testing the balanced embedding approach on D-Wave machines as well~\cite{bando2020probing}.

\begin{acknowledgments}
S.K. was supported by AFRL NYSTEC Contract (FA8750-19-3-6101). E.P. and
D.V. were supported by NASA Academic Mission Service contract NAMS
(NNA16BD14C), and by the Office of the Director of National Intelligence
(ODNI) and the Intelligence Advanced Research Projects Activity (IARPA),
via IAA 145483.
All authors acknowledge useful discussions with NASA QuAIL team members.
\end{acknowledgments}

\bibliography{paper}

\begin{thebibliography}{38}%
\makeatletter
\providecommand \@ifxundefined [1]{%
 \@ifx{#1\undefined}
}%
\providecommand \@ifnum [1]{%
 \ifnum #1\expandafter \@firstoftwo
 \else \expandafter \@secondoftwo
 \fi
}%
\providecommand \@ifx [1]{%
 \ifx #1\expandafter \@firstoftwo
 \else \expandafter \@secondoftwo
 \fi
}%
\providecommand \natexlab [1]{#1}%
\providecommand \enquote  [1]{``#1''}%
\providecommand \bibnamefont  [1]{#1}%
\providecommand \bibfnamefont [1]{#1}%
\providecommand \citenamefont [1]{#1}%
\providecommand \href@noop [0]{\@secondoftwo}%
\providecommand \href [0]{\begingroup \@sanitize@url \@href}%
\providecommand \@href[1]{\@@startlink{#1}\@@href}%
\providecommand \@@href[1]{\endgroup#1\@@endlink}%
\providecommand \@sanitize@url [0]{\catcode `\\12\catcode `\$12\catcode
  `\&12\catcode `\#12\catcode `\^12\catcode `\_12\catcode `\%12\relax}%
\providecommand \@@startlink[1]{}%
\providecommand \@@endlink[0]{}%
\providecommand \url  [0]{\begingroup\@sanitize@url \@url }%
\providecommand \@url [1]{\endgroup\@href {#1}{\urlprefix }}%
\providecommand \urlprefix  [0]{URL }%
\providecommand \Eprint [0]{\href }%
\providecommand \doibase [0]{http://dx.doi.org/}%
\providecommand \selectlanguage [0]{\@gobble}%
\providecommand \bibinfo  [0]{\@secondoftwo}%
\providecommand \bibfield  [0]{\@secondoftwo}%
\providecommand \translation [1]{[#1]}%
\providecommand \BibitemOpen [0]{}%
\providecommand \bibitemStop [0]{}%
\providecommand \bibitemNoStop [0]{.\EOS\space}%
\providecommand \EOS [0]{\spacefactor3000\relax}%
\providecommand \BibitemShut  [1]{\csname bibitem#1\endcsname}%
\let\auto@bib@innerbib\@empty
\bibitem [{\citenamefont {Kadowaki}\ and\ \citenamefont
  {Nishimori}(1998)}]{kadowaki1998quantum}%
  \BibitemOpen
  \bibfield  {author} {\bibinfo {author} {\bibfnamefont {T.}~\bibnamefont
  {Kadowaki}}\ and\ \bibinfo {author} {\bibfnamefont {H.}~\bibnamefont
  {Nishimori}},\ }\href@noop {} {\bibfield  {journal} {\bibinfo  {journal}
  {Physical Review E}\ }\textbf {\bibinfo {volume} {58}},\ \bibinfo {pages}
  {5355} (\bibinfo {year} {1998})}\BibitemShut {NoStop}%
\bibitem [{\citenamefont {Farhi}\ \emph {et~al.}()\citenamefont {Farhi},
  \citenamefont {Goldstone}, \citenamefont {Gutmann},\ and\ \citenamefont
  {Sipser}}]{farhi2000quantum}%
  \BibitemOpen
  \bibfield  {author} {\bibinfo {author} {\bibfnamefont {E.}~\bibnamefont
  {Farhi}}, \bibinfo {author} {\bibfnamefont {J.}~\bibnamefont {Goldstone}},
  \bibinfo {author} {\bibfnamefont {S.}~\bibnamefont {Gutmann}}, \ and\
  \bibinfo {author} {\bibfnamefont {M.}~\bibnamefont {Sipser}},\ }\href@noop {}
  {\ }\Eprint {http://arxiv.org/abs/quant-ph/0001106} {arXiv:quant-ph/0001106}
  \BibitemShut {NoStop}%
\bibitem [{\citenamefont {Farhi}\ \emph {et~al.}(2001)\citenamefont {Farhi},
  \citenamefont {Goldstone}, \citenamefont {Gutmann}, \citenamefont {Lapan},
  \citenamefont {Lundgren},\ and\ \citenamefont {Preda}}]{farhi2001quantum}%
  \BibitemOpen
  \bibfield  {author} {\bibinfo {author} {\bibfnamefont {E.}~\bibnamefont
  {Farhi}}, \bibinfo {author} {\bibfnamefont {J.}~\bibnamefont {Goldstone}},
  \bibinfo {author} {\bibfnamefont {S.}~\bibnamefont {Gutmann}}, \bibinfo
  {author} {\bibfnamefont {J.}~\bibnamefont {Lapan}}, \bibinfo {author}
  {\bibfnamefont {A.}~\bibnamefont {Lundgren}}, \ and\ \bibinfo {author}
  {\bibfnamefont {D.}~\bibnamefont {Preda}},\ }\href@noop {} {\bibfield
  {journal} {\bibinfo  {journal} {Science}\ }\textbf {\bibinfo {volume}
  {292}},\ \bibinfo {pages} {472} (\bibinfo {year} {2001})}\BibitemShut
  {NoStop}%
\bibitem [{\citenamefont {Das}\ and\ \citenamefont
  {Chakrabarti}(2008)}]{das2008colloquium}%
  \BibitemOpen
  \bibfield  {author} {\bibinfo {author} {\bibfnamefont {A.}~\bibnamefont
  {Das}}\ and\ \bibinfo {author} {\bibfnamefont {B.~K.}\ \bibnamefont
  {Chakrabarti}},\ }\href@noop {} {\bibfield  {journal} {\bibinfo  {journal}
  {Reviews of Modern Physics}\ }\textbf {\bibinfo {volume} {80}},\ \bibinfo
  {pages} {1061} (\bibinfo {year} {2008})}\BibitemShut {NoStop}%
\bibitem [{\citenamefont {Albash}\ and\ \citenamefont
  {Lidar}(2018)}]{albash2018adiabatic}%
  \BibitemOpen
  \bibfield  {author} {\bibinfo {author} {\bibfnamefont {T.}~\bibnamefont
  {Albash}}\ and\ \bibinfo {author} {\bibfnamefont {D.~A.}\ \bibnamefont
  {Lidar}},\ }\href@noop {} {\bibfield  {journal} {\bibinfo  {journal} {Reviews
  of Modern Physics}\ }\textbf {\bibinfo {volume} {90}},\ \bibinfo {pages}
  {015002} (\bibinfo {year} {2018})}\BibitemShut {NoStop}%
\bibitem [{\citenamefont {Choi}(2008)}]{choi2008minor}%
  \BibitemOpen
  \bibfield  {author} {\bibinfo {author} {\bibfnamefont {V.}~\bibnamefont
  {Choi}},\ }\href@noop {} {\bibfield  {journal} {\bibinfo  {journal} {Quantum
  Information Processing}\ }\textbf {\bibinfo {volume} {7}},\ \bibinfo {pages}
  {193} (\bibinfo {year} {2008})}\BibitemShut {NoStop}%
\bibitem [{\citenamefont {Choi}(2011)}]{choi2011minor}%
  \BibitemOpen
  \bibfield  {author} {\bibinfo {author} {\bibfnamefont {V.}~\bibnamefont
  {Choi}},\ }\href@noop {} {\bibfield  {journal} {\bibinfo  {journal} {Quantum
  Information Processing}\ }\textbf {\bibinfo {volume} {10}},\ \bibinfo {pages}
  {343} (\bibinfo {year} {2011})}\BibitemShut {NoStop}%
\bibitem [{\citenamefont {Hilton}\ \emph {et~al.}(2019)\citenamefont {Hilton},
  \citenamefont {Roy}, \citenamefont {Bunyk}, \citenamefont {King},
  \citenamefont {Boothby}, \citenamefont {Harris},\ and\ \citenamefont
  {Deng}}]{hilton2019systems}%
  \BibitemOpen
  \bibfield  {author} {\bibinfo {author} {\bibfnamefont {J.~P.}\ \bibnamefont
  {Hilton}}, \bibinfo {author} {\bibfnamefont {A.~P.}\ \bibnamefont {Roy}},
  \bibinfo {author} {\bibfnamefont {P.~I.}\ \bibnamefont {Bunyk}}, \bibinfo
  {author} {\bibfnamefont {A.~D.}\ \bibnamefont {King}}, \bibinfo {author}
  {\bibfnamefont {K.~T.}\ \bibnamefont {Boothby}}, \bibinfo {author}
  {\bibfnamefont {R.~G.}\ \bibnamefont {Harris}}, \ and\ \bibinfo {author}
  {\bibfnamefont {C.}~\bibnamefont {Deng}},\ }\href@noop {} {\enquote {\bibinfo
  {title} {Systems and methods for increasing analog processor connectivity},}\
  }\bibinfo {howpublished} {US Patent 10,268,622} (\bibinfo {year}
  {2019})\BibitemShut {NoStop}%
\bibitem [{\citenamefont {King}\ and\ \citenamefont
  {McGeoch}()}]{king2014algorithm}%
  \BibitemOpen
  \bibfield  {author} {\bibinfo {author} {\bibfnamefont {A.~D.}\ \bibnamefont
  {King}}\ and\ \bibinfo {author} {\bibfnamefont {C.~C.}\ \bibnamefont
  {McGeoch}},\ }\href@noop {} {\ }\Eprint {http://arxiv.org/abs/1410.2628}
  {arXiv:1410.2628} \BibitemShut {NoStop}%
\bibitem [{\citenamefont {Venturelli}\ \emph {et~al.}(2015)\citenamefont
  {Venturelli}, \citenamefont {Mandra}, \citenamefont {Knysh}, \citenamefont
  {O'Gorman}, \citenamefont {Biswas},\ and\ \citenamefont
  {Smelyanskiy}}]{venturelli2015quantum}%
  \BibitemOpen
  \bibfield  {author} {\bibinfo {author} {\bibfnamefont {D.}~\bibnamefont
  {Venturelli}}, \bibinfo {author} {\bibfnamefont {S.}~\bibnamefont {Mandra}},
  \bibinfo {author} {\bibfnamefont {S.}~\bibnamefont {Knysh}}, \bibinfo
  {author} {\bibfnamefont {B.}~\bibnamefont {O'Gorman}}, \bibinfo {author}
  {\bibfnamefont {R.}~\bibnamefont {Biswas}}, \ and\ \bibinfo {author}
  {\bibfnamefont {V.}~\bibnamefont {Smelyanskiy}},\ }\href@noop {} {\bibfield
  {journal} {\bibinfo  {journal} {Physical Review X}\ }\textbf {\bibinfo
  {volume} {5}},\ \bibinfo {pages} {031040} (\bibinfo {year}
  {2015})}\BibitemShut {NoStop}%
\bibitem [{\citenamefont {Fang}\ and\ \citenamefont
  {Warburton}()}]{fang2019minimizing}%
  \BibitemOpen
  \bibfield  {author} {\bibinfo {author} {\bibfnamefont {Y.-L.}\ \bibnamefont
  {Fang}}\ and\ \bibinfo {author} {\bibfnamefont {P.}~\bibnamefont
  {Warburton}},\ }\href@noop {} {\ }\Eprint {http://arxiv.org/abs/1905.03291}
  {arXiv:1905.03291} \BibitemShut {NoStop}%
\bibitem [{\citenamefont {Harris}\ \emph {et~al.}(2018)\citenamefont {Harris},
  \citenamefont {Sato}, \citenamefont {Berkley}, \citenamefont {Reis},
  \citenamefont {Altomare}, \citenamefont {Amin}, \citenamefont {Boothby},
  \citenamefont {Bunyk}, \citenamefont {Deng}, \citenamefont {Enderud} \emph
  {et~al.}}]{harris2018phase}%
  \BibitemOpen
  \bibfield  {author} {\bibinfo {author} {\bibfnamefont {R.}~\bibnamefont
  {Harris}}, \bibinfo {author} {\bibfnamefont {Y.}~\bibnamefont {Sato}},
  \bibinfo {author} {\bibfnamefont {A.}~\bibnamefont {Berkley}}, \bibinfo
  {author} {\bibfnamefont {M.}~\bibnamefont {Reis}}, \bibinfo {author}
  {\bibfnamefont {F.}~\bibnamefont {Altomare}}, \bibinfo {author}
  {\bibfnamefont {M.}~\bibnamefont {Amin}}, \bibinfo {author} {\bibfnamefont
  {K.}~\bibnamefont {Boothby}}, \bibinfo {author} {\bibfnamefont
  {P.}~\bibnamefont {Bunyk}}, \bibinfo {author} {\bibfnamefont
  {C.}~\bibnamefont {Deng}}, \bibinfo {author} {\bibfnamefont {C.}~\bibnamefont
  {Enderud}},  \emph {et~al.},\ }\href@noop {} {\bibfield  {journal} {\bibinfo
  {journal} {Science}\ }\textbf {\bibinfo {volume} {361}},\ \bibinfo {pages}
  {162} (\bibinfo {year} {2018})}\BibitemShut {NoStop}%
\bibitem [{\citenamefont {Hastings}()}]{hastings2020power}%
  \BibitemOpen
  \bibfield  {author} {\bibinfo {author} {\bibfnamefont {M.~B.}\ \bibnamefont
  {Hastings}},\ }\href@noop {} {\ }\Eprint {http://arxiv.org/abs/2005.03791}
  {arXiv:2005.03791} \BibitemShut {NoStop}%
\bibitem [{\citenamefont {Zurek}\ \emph {et~al.}(2005)\citenamefont {Zurek},
  \citenamefont {Dorner},\ and\ \citenamefont {Zoller}}]{zurek2005dynamics}%
  \BibitemOpen
  \bibfield  {author} {\bibinfo {author} {\bibfnamefont {W.~H.}\ \bibnamefont
  {Zurek}}, \bibinfo {author} {\bibfnamefont {U.}~\bibnamefont {Dorner}}, \
  and\ \bibinfo {author} {\bibfnamefont {P.}~\bibnamefont {Zoller}},\
  }\href@noop {} {\bibfield  {journal} {\bibinfo  {journal} {Physical Review
  Letters}\ }\textbf {\bibinfo {volume} {95}},\ \bibinfo {pages} {105701}
  (\bibinfo {year} {2005})}\BibitemShut {NoStop}%
\bibitem [{\citenamefont {Dziarmaga}(2005)}]{dziarmaga2005dynamics}%
  \BibitemOpen
  \bibfield  {author} {\bibinfo {author} {\bibfnamefont {J.}~\bibnamefont
  {Dziarmaga}},\ }\href@noop {} {\bibfield  {journal} {\bibinfo  {journal}
  {Physical Review Letters}\ }\textbf {\bibinfo {volume} {95}},\ \bibinfo
  {pages} {245701} (\bibinfo {year} {2005})}\BibitemShut {NoStop}%
\bibitem [{\citenamefont {Fisher}(1992)}]{fisher1992random}%
  \BibitemOpen
  \bibfield  {author} {\bibinfo {author} {\bibfnamefont {D.~S.}\ \bibnamefont
  {Fisher}},\ }\href@noop {} {\bibfield  {journal} {\bibinfo  {journal}
  {Physical Review Letters}\ }\textbf {\bibinfo {volume} {69}},\ \bibinfo
  {pages} {534} (\bibinfo {year} {1992})}\BibitemShut {NoStop}%
\bibitem [{\citenamefont {Fisher}(1995)}]{fisher1995critical}%
  \BibitemOpen
  \bibfield  {author} {\bibinfo {author} {\bibfnamefont {D.~S.}\ \bibnamefont
  {Fisher}},\ }\href@noop {} {\bibfield  {journal} {\bibinfo  {journal}
  {Physical Review B}\ }\textbf {\bibinfo {volume} {51}},\ \bibinfo {pages}
  {6411} (\bibinfo {year} {1995})}\BibitemShut {NoStop}%
\bibitem [{\citenamefont {Griffiths}(1969)}]{griffiths1969nonanalytic}%
  \BibitemOpen
  \bibfield  {author} {\bibinfo {author} {\bibfnamefont {R.~B.}\ \bibnamefont
  {Griffiths}},\ }\href@noop {} {\bibfield  {journal} {\bibinfo  {journal}
  {Physical Review Letters}\ }\textbf {\bibinfo {volume} {23}},\ \bibinfo
  {pages} {17} (\bibinfo {year} {1969})}\BibitemShut {NoStop}%
\bibitem [{\citenamefont {McCoy}(1969)}]{mccoy1969incompleteness}%
  \BibitemOpen
  \bibfield  {author} {\bibinfo {author} {\bibfnamefont {B.~M.}\ \bibnamefont
  {McCoy}},\ }\href@noop {} {\bibfield  {journal} {\bibinfo  {journal}
  {Physical Review Letters}\ }\textbf {\bibinfo {volume} {23}},\ \bibinfo
  {pages} {383} (\bibinfo {year} {1969})}\BibitemShut {NoStop}%
\bibitem [{\citenamefont {Pudenz}\ \emph {et~al.}(2014)\citenamefont {Pudenz},
  \citenamefont {Albash},\ and\ \citenamefont {Lidar}}]{pudenz2014error}%
  \BibitemOpen
  \bibfield  {author} {\bibinfo {author} {\bibfnamefont {K.~L.}\ \bibnamefont
  {Pudenz}}, \bibinfo {author} {\bibfnamefont {T.}~\bibnamefont {Albash}}, \
  and\ \bibinfo {author} {\bibfnamefont {D.~A.}\ \bibnamefont {Lidar}},\
  }\href@noop {} {\bibfield  {journal} {\bibinfo  {journal} {Nature
  Communications}\ }\textbf {\bibinfo {volume} {5}},\ \bibinfo {pages} {1}
  (\bibinfo {year} {2014})}\BibitemShut {NoStop}%
\bibitem [{\citenamefont {Pudenz}\ \emph {et~al.}(2015)\citenamefont {Pudenz},
  \citenamefont {Albash},\ and\ \citenamefont {Lidar}}]{pudenz2015quantum}%
  \BibitemOpen
  \bibfield  {author} {\bibinfo {author} {\bibfnamefont {K.~L.}\ \bibnamefont
  {Pudenz}}, \bibinfo {author} {\bibfnamefont {T.}~\bibnamefont {Albash}}, \
  and\ \bibinfo {author} {\bibfnamefont {D.~A.}\ \bibnamefont {Lidar}},\
  }\href@noop {} {\bibfield  {journal} {\bibinfo  {journal} {Physical Review
  A}\ }\textbf {\bibinfo {volume} {91}},\ \bibinfo {pages} {042302} (\bibinfo
  {year} {2015})}\BibitemShut {NoStop}%
\bibitem [{\citenamefont {Vinci}\ \emph {et~al.}(2015)\citenamefont {Vinci},
  \citenamefont {Albash}, \citenamefont {Paz-Silva}, \citenamefont {Hen},\ and\
  \citenamefont {Lidar}}]{vinci2015quantum}%
  \BibitemOpen
  \bibfield  {author} {\bibinfo {author} {\bibfnamefont {W.}~\bibnamefont
  {Vinci}}, \bibinfo {author} {\bibfnamefont {T.}~\bibnamefont {Albash}},
  \bibinfo {author} {\bibfnamefont {G.}~\bibnamefont {Paz-Silva}}, \bibinfo
  {author} {\bibfnamefont {I.}~\bibnamefont {Hen}}, \ and\ \bibinfo {author}
  {\bibfnamefont {D.~A.}\ \bibnamefont {Lidar}},\ }\href@noop {} {\bibfield
  {journal} {\bibinfo  {journal} {Physical Review A}\ }\textbf {\bibinfo
  {volume} {92}},\ \bibinfo {pages} {042310} (\bibinfo {year}
  {2015})}\BibitemShut {NoStop}%
\bibitem [{\citenamefont {Matsuura}\ \emph {et~al.}(2017)\citenamefont
  {Matsuura}, \citenamefont {Nishimori}, \citenamefont {Vinci}, \citenamefont
  {Albash},\ and\ \citenamefont {Lidar}}]{matsuura2017quantum}%
  \BibitemOpen
  \bibfield  {author} {\bibinfo {author} {\bibfnamefont {S.}~\bibnamefont
  {Matsuura}}, \bibinfo {author} {\bibfnamefont {H.}~\bibnamefont {Nishimori}},
  \bibinfo {author} {\bibfnamefont {W.}~\bibnamefont {Vinci}}, \bibinfo
  {author} {\bibfnamefont {T.}~\bibnamefont {Albash}}, \ and\ \bibinfo {author}
  {\bibfnamefont {D.~A.}\ \bibnamefont {Lidar}},\ }\href@noop {} {\bibfield
  {journal} {\bibinfo  {journal} {Physical Review A}\ }\textbf {\bibinfo
  {volume} {95}},\ \bibinfo {pages} {022308} (\bibinfo {year}
  {2017})}\BibitemShut {NoStop}%
\bibitem [{\citenamefont {Lucas}(2014)}]{lucas2014ising}%
  \BibitemOpen
  \bibfield  {author} {\bibinfo {author} {\bibfnamefont {A.}~\bibnamefont
  {Lucas}},\ }\href@noop {} {\bibfield  {journal} {\bibinfo  {journal}
  {Frontiers in Physics}\ }\textbf {\bibinfo {volume} {2}},\ \bibinfo {pages}
  {5} (\bibinfo {year} {2014})}\BibitemShut {NoStop}%
\bibitem [{\citenamefont {Dziarmaga}(2006)}]{dziarmaga2006dynamics}%
  \BibitemOpen
  \bibfield  {author} {\bibinfo {author} {\bibfnamefont {J.}~\bibnamefont
  {Dziarmaga}},\ }\href@noop {} {\bibfield  {journal} {\bibinfo  {journal}
  {Physical Review B}\ }\textbf {\bibinfo {volume} {74}},\ \bibinfo {pages}
  {064416} (\bibinfo {year} {2006})}\BibitemShut {NoStop}%
\bibitem [{\citenamefont {Caneva}\ \emph {et~al.}(2007)\citenamefont {Caneva},
  \citenamefont {Fazio},\ and\ \citenamefont {Santoro}}]{caneva2007adiabatic}%
  \BibitemOpen
  \bibfield  {author} {\bibinfo {author} {\bibfnamefont {T.}~\bibnamefont
  {Caneva}}, \bibinfo {author} {\bibfnamefont {R.}~\bibnamefont {Fazio}}, \
  and\ \bibinfo {author} {\bibfnamefont {G.~E.}\ \bibnamefont {Santoro}},\
  }\href@noop {} {\bibfield  {journal} {\bibinfo  {journal} {Physical Review
  B}\ }\textbf {\bibinfo {volume} {76}},\ \bibinfo {pages} {144427} (\bibinfo
  {year} {2007})}\BibitemShut {NoStop}%
\bibitem [{\citenamefont {Fradkin}(2013)}]{fradkin2013field}%
  \BibitemOpen
  \bibfield  {author} {\bibinfo {author} {\bibfnamefont {E.}~\bibnamefont
  {Fradkin}},\ }\href@noop {} {\emph {\bibinfo {title} {Field theories of
  condensed matter physics}}}\ (\bibinfo  {publisher} {Cambridge University
  Press},\ \bibinfo {year} {2013})\BibitemShut {NoStop}%
\bibitem [{\citenamefont {Blanes}\ \emph {et~al.}(2000)\citenamefont {Blanes},
  \citenamefont {Casas},\ and\ \citenamefont {Ros}}]{blanes2000improved}%
  \BibitemOpen
  \bibfield  {author} {\bibinfo {author} {\bibfnamefont {S.}~\bibnamefont
  {Blanes}}, \bibinfo {author} {\bibfnamefont {F.}~\bibnamefont {Casas}}, \
  and\ \bibinfo {author} {\bibfnamefont {J.}~\bibnamefont {Ros}},\ }\href@noop
  {} {\bibfield  {journal} {\bibinfo  {journal} {BIT Numerical Mathematics}\
  }\textbf {\bibinfo {volume} {40}},\ \bibinfo {pages} {434} (\bibinfo {year}
  {2000})}\BibitemShut {NoStop}%
\bibitem [{\citenamefont {Blanes}\ \emph {et~al.}(2002)\citenamefont {Blanes},
  \citenamefont {Casas},\ and\ \citenamefont {Ros}}]{blanes2002high}%
  \BibitemOpen
  \bibfield  {author} {\bibinfo {author} {\bibfnamefont {S.}~\bibnamefont
  {Blanes}}, \bibinfo {author} {\bibfnamefont {F.}~\bibnamefont {Casas}}, \
  and\ \bibinfo {author} {\bibfnamefont {J.}~\bibnamefont {Ros}},\ }\href@noop
  {} {\bibfield  {journal} {\bibinfo  {journal} {BIT Numerical Mathematics}\
  }\textbf {\bibinfo {volume} {42}},\ \bibinfo {pages} {262} (\bibinfo {year}
  {2002})}\BibitemShut {NoStop}%
\bibitem [{\citenamefont {Mbeng}\ \emph {et~al.}(2019)\citenamefont {Mbeng},
  \citenamefont {Privitera}, \citenamefont {Arceci},\ and\ \citenamefont
  {Santoro}}]{mbeng2019dynamics}%
  \BibitemOpen
  \bibfield  {author} {\bibinfo {author} {\bibfnamefont {G.~B.}\ \bibnamefont
  {Mbeng}}, \bibinfo {author} {\bibfnamefont {L.}~\bibnamefont {Privitera}},
  \bibinfo {author} {\bibfnamefont {L.}~\bibnamefont {Arceci}}, \ and\ \bibinfo
  {author} {\bibfnamefont {G.~E.}\ \bibnamefont {Santoro}},\ }\href@noop {}
  {\bibfield  {journal} {\bibinfo  {journal} {Physical Review B}\ }\textbf
  {\bibinfo {volume} {99}},\ \bibinfo {pages} {064201} (\bibinfo {year}
  {2019})}\BibitemShut {NoStop}%
\bibitem [{\citenamefont {Dormand}\ and\ \citenamefont
  {Prince}(1980)}]{dormand1980family}%
  \BibitemOpen
  \bibfield  {author} {\bibinfo {author} {\bibfnamefont {J.~R.}\ \bibnamefont
  {Dormand}}\ and\ \bibinfo {author} {\bibfnamefont {P.~J.}\ \bibnamefont
  {Prince}},\ }\href@noop {} {\bibfield  {journal} {\bibinfo  {journal}
  {Journal of Computational and Applied Mathematics}\ }\textbf {\bibinfo
  {volume} {6}},\ \bibinfo {pages} {19} (\bibinfo {year} {1980})}\BibitemShut
  {NoStop}%
\bibitem [{\citenamefont {Rams}\ \emph {et~al.}(2016)\citenamefont {Rams},
  \citenamefont {Mohseni},\ and\ \citenamefont {del
  Campo}}]{rams2016inhomogeneous}%
  \BibitemOpen
  \bibfield  {author} {\bibinfo {author} {\bibfnamefont {M.~M.}\ \bibnamefont
  {Rams}}, \bibinfo {author} {\bibfnamefont {M.}~\bibnamefont {Mohseni}}, \
  and\ \bibinfo {author} {\bibfnamefont {A.}~\bibnamefont {del Campo}},\
  }\href@noop {} {\bibfield  {journal} {\bibinfo  {journal} {New Journal of
  Physics}\ }\textbf {\bibinfo {volume} {18}},\ \bibinfo {pages} {123034}
  (\bibinfo {year} {2016})}\BibitemShut {NoStop}%
\bibitem [{\citenamefont {Knysh}(2016)}]{knysh2016zero}%
  \BibitemOpen
  \bibfield  {author} {\bibinfo {author} {\bibfnamefont {S.}~\bibnamefont
  {Knysh}},\ }\href@noop {} {\bibfield  {journal} {\bibinfo  {journal} {Nature
  Communications}\ }\textbf {\bibinfo {volume} {7}},\ \bibinfo {pages} {1}
  (\bibinfo {year} {2016})}\BibitemShut {NoStop}%
\bibitem [{\citenamefont {Roberts}\ \emph {et~al.}(2020)\citenamefont
  {Roberts}, \citenamefont {Cincio}, \citenamefont {Saxena}, \citenamefont
  {Petukhov},\ and\ \citenamefont {Knysh}}]{roberts2020noise}%
  \BibitemOpen
  \bibfield  {author} {\bibinfo {author} {\bibfnamefont {D.}~\bibnamefont
  {Roberts}}, \bibinfo {author} {\bibfnamefont {L.}~\bibnamefont {Cincio}},
  \bibinfo {author} {\bibfnamefont {A.}~\bibnamefont {Saxena}}, \bibinfo
  {author} {\bibfnamefont {A.}~\bibnamefont {Petukhov}}, \ and\ \bibinfo
  {author} {\bibfnamefont {S.}~\bibnamefont {Knysh}},\ }\href@noop {}
  {\bibfield  {journal} {\bibinfo  {journal} {Physical Review A}\ }\textbf
  {\bibinfo {volume} {101}},\ \bibinfo {pages} {042317} (\bibinfo {year}
  {2020})}\BibitemShut {NoStop}%
\bibitem [{\citenamefont {Novikov}\ \emph {et~al.}(2018)\citenamefont
  {Novikov}, \citenamefont {Hinkey}, \citenamefont {Disseler}, \citenamefont
  {Basham}, \citenamefont {Albash}, \citenamefont {Risinger}, \citenamefont
  {Ferguson}, \citenamefont {Lidar},\ and\ \citenamefont
  {Zick}}]{novikov2018exploring}%
  \BibitemOpen
  \bibfield  {author} {\bibinfo {author} {\bibfnamefont {S.}~\bibnamefont
  {Novikov}}, \bibinfo {author} {\bibfnamefont {R.}~\bibnamefont {Hinkey}},
  \bibinfo {author} {\bibfnamefont {S.}~\bibnamefont {Disseler}}, \bibinfo
  {author} {\bibfnamefont {J.~I.}\ \bibnamefont {Basham}}, \bibinfo {author}
  {\bibfnamefont {T.}~\bibnamefont {Albash}}, \bibinfo {author} {\bibfnamefont
  {A.}~\bibnamefont {Risinger}}, \bibinfo {author} {\bibfnamefont
  {D.}~\bibnamefont {Ferguson}}, \bibinfo {author} {\bibfnamefont {D.~A.}\
  \bibnamefont {Lidar}}, \ and\ \bibinfo {author} {\bibfnamefont {K.~M.}\
  \bibnamefont {Zick}},\ }in\ \href@noop {} {\emph {\bibinfo {booktitle} {2018
  IEEE International Conference on Rebooting Computing (ICRC)}}}\ (\bibinfo
  {organization} {IEEE},\ \bibinfo {year} {2018})\ pp.\ \bibinfo {pages}
  {1--7}\BibitemShut {NoStop}%
\bibitem [{\citenamefont {Smelyanskiy}\ \emph {et~al.}(2017)\citenamefont
  {Smelyanskiy}, \citenamefont {Venturelli}, \citenamefont {Perdomo-Ortiz},
  \citenamefont {Knysh},\ and\ \citenamefont
  {Dykman}}]{smelyanskiy2017quantum}%
  \BibitemOpen
  \bibfield  {author} {\bibinfo {author} {\bibfnamefont {V.~N.}\ \bibnamefont
  {Smelyanskiy}}, \bibinfo {author} {\bibfnamefont {D.}~\bibnamefont
  {Venturelli}}, \bibinfo {author} {\bibfnamefont {A.}~\bibnamefont
  {Perdomo-Ortiz}}, \bibinfo {author} {\bibfnamefont {S.}~\bibnamefont
  {Knysh}}, \ and\ \bibinfo {author} {\bibfnamefont {M.~I.}\ \bibnamefont
  {Dykman}},\ }\href@noop {} {\bibfield  {journal} {\bibinfo  {journal}
  {Physical Review Letters}\ }\textbf {\bibinfo {volume} {118}},\ \bibinfo
  {pages} {066802} (\bibinfo {year} {2017})}\BibitemShut {NoStop}%
\bibitem [{\citenamefont {Mishra}\ \emph {et~al.}(2018)\citenamefont {Mishra},
  \citenamefont {Albash},\ and\ \citenamefont {Lidar}}]{mishra2018finite}%
  \BibitemOpen
  \bibfield  {author} {\bibinfo {author} {\bibfnamefont {A.}~\bibnamefont
  {Mishra}}, \bibinfo {author} {\bibfnamefont {T.}~\bibnamefont {Albash}}, \
  and\ \bibinfo {author} {\bibfnamefont {D.~A.}\ \bibnamefont {Lidar}},\
  }\href@noop {} {\bibfield  {journal} {\bibinfo  {journal} {Nature
  Communications}\ }\textbf {\bibinfo {volume} {9}},\ \bibinfo {pages} {1}
  (\bibinfo {year} {2018})}\BibitemShut {NoStop}%
\bibitem [{\citenamefont {Bando}\ \emph {et~al.}()\citenamefont {Bando},
  \citenamefont {Susa}, \citenamefont {Oshiyama}, \citenamefont {Shibata},
  \citenamefont {Ohzeki}, \citenamefont {G{\'o}mez-Ruiz}, \citenamefont
  {Lidar}, \citenamefont {del Campo}, \citenamefont {Suzuki},\ and\
  \citenamefont {Nishimori}}]{bando2020probing}%
  \BibitemOpen
  \bibfield  {author} {\bibinfo {author} {\bibfnamefont {Y.}~\bibnamefont
  {Bando}}, \bibinfo {author} {\bibfnamefont {Y.}~\bibnamefont {Susa}},
  \bibinfo {author} {\bibfnamefont {H.}~\bibnamefont {Oshiyama}}, \bibinfo
  {author} {\bibfnamefont {N.}~\bibnamefont {Shibata}}, \bibinfo {author}
  {\bibfnamefont {M.}~\bibnamefont {Ohzeki}}, \bibinfo {author} {\bibfnamefont
  {F.~J.}\ \bibnamefont {G{\'o}mez-Ruiz}}, \bibinfo {author} {\bibfnamefont
  {D.~A.}\ \bibnamefont {Lidar}}, \bibinfo {author} {\bibfnamefont
  {A.}~\bibnamefont {del Campo}}, \bibinfo {author} {\bibfnamefont
  {S.}~\bibnamefont {Suzuki}}, \ and\ \bibinfo {author} {\bibfnamefont
  {H.}~\bibnamefont {Nishimori}},\ }\href@noop {} {\ }\Eprint
  {http://arxiv.org/abs/2001.11637} {arXiv:2001.11637} \BibitemShut {NoStop}%
\end{thebibliography}%

\newpage
\appendix
\onecolumngrid

\section{\label{app:Eigenspectrum}Eigenspectrum}

Diagonalization of one-dimensional Ising model is performed using
Jordan-Wigner transformation. The Hamiltonian reads in spin representation as
\begin{equation}
  H = - \sum_{i = 1}^{N - 1} J_i X_{i - 1} X_i + \Gamma \sum_{i = 0}^{N - 1}
  Z_i,
\end{equation}
where we rotated the basis around $y$-axis by $\pi / 2$. We introduce $2 N$
Majorana fermions defined via
\begin{equation}
  \chi_{2 i} = X_i \prod_{k = 0}^{i - 1} Z_k, \qquad \chi_{2 i + 1} = Y_i
  \prod_{k = 0}^{i - 1} Z_k,
\end{equation}
and obeying anti-commutation relations $\{ \chi_i, \chi_j \} = 2 \delta_{i j}$
as can be straightforwardly verified. In terms of these, the Ising Hamiltonian
can be rewritten as
\begin{equation}
  H = - \ii \sum_{i = 1}^{N - 1} J_i \chi_{2 i - 1} \chi_{2 i} + \ii
  \Gamma \sum_{i = 0}^{N - 1} \chi_{2 i} \chi_{2 i + 1} . \label{HMaj}
\end{equation}
The Hamiltonian can be equivalently represented in terms of Dirac
quasiparticles
\begin{equation}
  H = \sum_{\nu} \epsilon_{\nu} (\gamma_{\nu}^{\dag} \gamma_{\nu} -
  \gamma_{\nu}^{\dag} \gamma_{\nu}) .
\end{equation}
From this representation one obtains all $2^N$ energy levels:
\begin{equation}
  E_{\mathbf{n}} = E_0 + 2 \sum_{\nu} n_{\nu} \epsilon_{\nu}, \text{ where
  $n_{\nu} \in \{ 0, 1 \}$ are the occupation numbers and } E_0 = - \sum_{\nu}
  \epsilon_{\nu} .
\end{equation}
Using identities
\begin{equation}
  [\gamma_{\mu}, H] = 2 \epsilon_{\mu} \gamma_{\mu} \qquad \text{and} \qquad
  [\gamma_{\mu}^{\dag}, H] = - 2 \epsilon_{\mu} \gamma_{\mu}^{\dag}
\end{equation}
one obtains the single-fermion excitation energies and creation/annihilation
operators from the spectrum of the tridiagonal matrix
\begin{equation}
  \mathbf{M}= \left(\begin{array}{cccccccc}
    0 & \Gamma &  &  &  &  &  & \\
    -\Gamma & 0 & J_1 &  &  &  &  & \\
    & -J_1 & 0 & \Gamma &  &  &  & \\
    &  & -\Gamma & 0 & J_2 &  &  & \\
    &  &  & -J_2 & \ddots & \ddots &  & \\
    &  &  &  & \ddots & \ddots & J_{N - 1} & \\
    &  &  &  &  & -J_{N - 1} & 0 & \Gamma\\
    &  &  &  &  &  & -\Gamma & 0
  \end{array}\right)
\end{equation}
that appears Majorana representation of Eq. (\ref{HMaj}); that is $H =
\frac{1}{2} \ii \boldsymbol{\chi}^T \mathbf{M}\boldsymbol{\chi}$ where
$\boldsymbol{\chi}$ is a column vector of Majorana fermions $\chi_0, \chi_1,
\ldots, \chi_{2N-1}$.

To study the behavior of the gap we write the characteristic equation:
\begin{equation}
  \det (\lambda +\mathbf{M}^2) = 0, \quad \mathbf{M}^2 =
   - \left(\begin{array}{cccccccc}
     \Gamma^2 & - \Gamma J_1 &  &  &  &  &  & \\
     - \Gamma J_1 & \Gamma^2 + J_1^2 & \ddots &  &  &  &  & \\
     & \ddots & \ddots & - \Gamma J_{N-1} &  &  &  & \\
     &  & - \Gamma J_{N - 1} & \Gamma^2 + J_{N-1}^2 &  &  &  & \\
     &  &  &  & \Gamma^2 + J_1^2 & - \Gamma J_1 &  & \\
     &  &  &  & - \Gamma J_1 & \ddots & \ddots & \\
     &  &  &  &  & \ddots & \Gamma^2 + J_{N - 1}^2 & - \Gamma J_{N-1}\\
     &  &  &  &  &  & - \Gamma J_{N - 1} & \Gamma^2
   \end{array}\right).
\end{equation}
Above we rearranged the rows and columns of $\mathbf{M}^2$ (by grouping
even-numbered and odd-numbered) in block-diagonal form. It is straightforward
to verify that both blocks yield the same characteristic equation (the terms
can be evaluated by computing minors via induction over $N$):
\begin{equation}
  \Gamma^{2N} - \lambda \Gamma^{2(N-1)} \sum_{0<k_1<k_2<N}
  \frac{\prod_{i=k_1}^{k_2-1} J_i^2}{\Gamma^{2(k_2-k_1)}} + \Gamma^{2(N-2)} \lambda^2 \sum_{0<k_1<k_2<k_3<k_4<N} \frac{\prod_{i=k_1}^{k_2-1} J_i^2}{\Gamma^{2(k_2-k_1)}} 
  \cdot \frac{\prod_{i=k_3}^{k_4-1} J_i^2}{\Gamma^{2(k_4 - k_3)}} 
  + \cdots = 0. \label{char}
\end{equation}
Indeed, the roots of characteristic equation should be doubly degenerate
corresponding to the pairs of eigenvalues $\pm \ii \epsilon_k$ of matrix
$\mathbf{M}$: $\lambda_k = \epsilon_k^2$.

In the region of interest of the disordered problem we expect that
$\epsilon_0 \ll \epsilon_1 \ll \epsilon_2 \ll \cdots$ and so on. With this in
mind, it suffices to retain terms up to $\lambda^2$ to estimate the gap to the
second excited state: we approximate $\epsilon_0$ and $\epsilon_1$ by
solutions to a quadratic equation (\ref{char}). To illustrate this in more
detail, Fig.~\ref{fig:cumJ} below presents plots of a function (defined for integer $1 \leqslant k \leqslant N$)
\begin{equation}
  W(k) = \sum_{i=1}^{k-1} \ln J_i - k \ln \Gamma
  \label{Wk}
\end{equation}
for three different values of $\Gamma$. On large scales $W(k)$ describes a
Wiener process (Brownian motion) so that it typically scales as $O\left(\sqrt{N}\right)$. The partial products inside the sums in Eq. (\ref{char}) may be expressed as $\ee^{2[W(k_2)-W (k_1)]}$ 
for a term linear in $\lambda$ 
or $\ee^{2[W(k_4)-W(k_3)+W(k_2)-W(k_1)]}$ for a term
quadratic in $\lambda$. Large exponent ensures that the sums are dominated by
just a few terms. Using $W_1$, $W_2$, $W_3$, and $W_4$ do denote the extremal
values as depicted in Fig.~\ref{fig:cumJ} we can write Vieta's formulae:
\begin{equation}
  \frac{1}{\lambda_0 \lambda_1} \sim \ee^{2(W_4-W_3+W_2-W_1)},
  \qquad \frac{1}{\lambda_0} + \frac{1}{\lambda_1} \sim
  \begin{cases}
    \ee^{2(W_4-W_1)} & \text{for } \Gamma < \Gamma_{\ast},\\
    \ee^{2(W_2-W_1)} & \text{for } \Gamma > \Gamma_{\ast},
  \end{cases}
\end{equation}
where $\Gamma_{\ast}$ has been chose so that $W_2 \approx W_4$. From this we
conclude that
\begin{align}
\epsilon_0 \sim \ee^{-(W_4-W_1)} \quad &\text{and} \quad \epsilon_1
   \sim \ee^{-(W_2-W_3)} && (\Gamma < \Gamma_{\ast}), \\
\epsilon_0 \sim \ee^{-(W_2-W_1)} \quad &\text{and} \quad \epsilon_1
   \sim \ee^{-(W_4-W_3)} && (\Gamma > \Gamma_{\ast}).
\end{align}

\begin{figure}[!ht]
  \includegraphics[width=\linewidth]{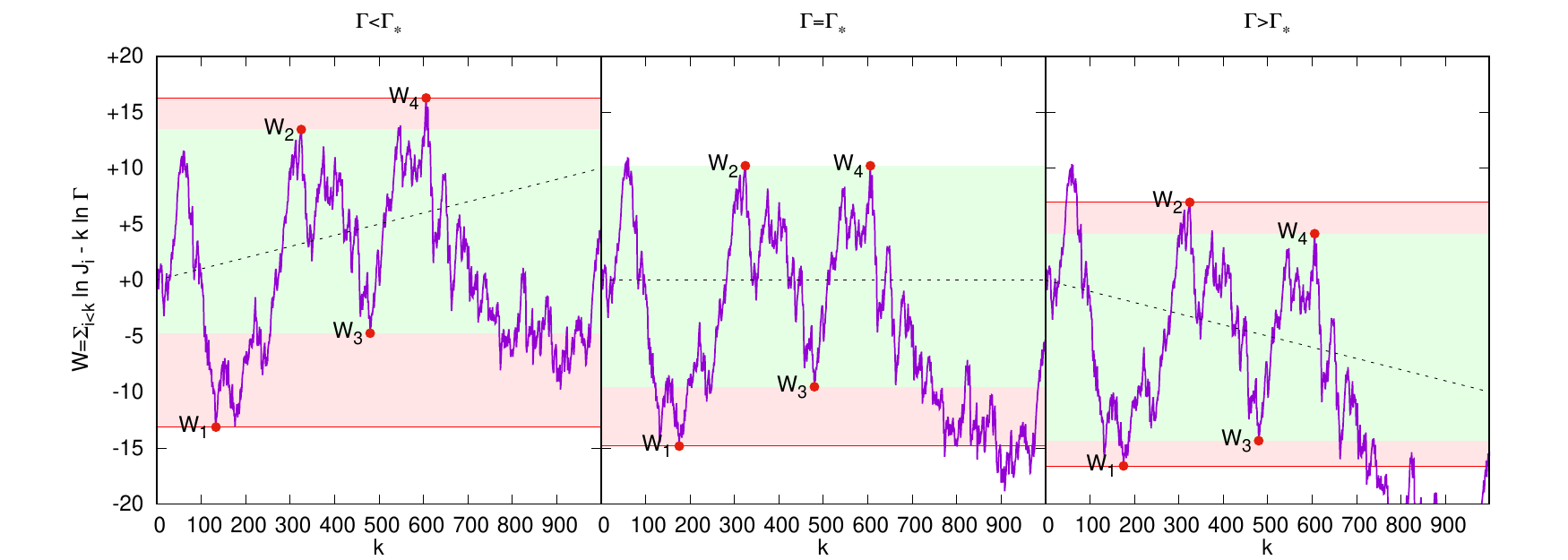}
  \caption{\label{fig:cumJ}
    Blue solid line shows realization of random process described 
    by Eq.~\ref{Wk}. Black dotted line, $-k (\ln \Gamma-\ln \Gamma_\ast)$, serves as visual guide to describe vertical shear as $\Gamma$ is varied. From left to right, subplots
    depicts scenarios $\Gamma<\Gamma_\ast$, $\Gamma=\Gamma_\ast$,
    and $\Gamma>\Gamma_\ast$ respectively. For each subplot, single-particle energy $\epsilon_0$ is given by the negative exponential of the height of the shaded red region and $\epsilon_1$ is the negative exponential of the height of shaded green region (which is entirely within the red region so that $\epsilon_0<\epsilon_1$). The height of the red region decreases from left to right whereas the height of the green region is largest at the center so that $\epsilon_1$ is minimized for $\Gamma \approx \Gamma_\ast$.
  }
\end{figure}

Changing $\Gamma$ applies a shear in vertical direction; while $\epsilon_0$ is
monotonic as a function of $\Gamma$, it is straightforward to see that
$\epsilon_1$ increases as $\Gamma$ moves away from $\Gamma_{\ast}$ in either
direction. Therefore $\Gamma = \Gamma_{\ast}$ is the approximate location of
the minimum gap.

The level of detail presented above is not necessary to establish the
stretched exponential scaling of the minimum gap $E_2-E_0 \sim \ee^{-c\sqrt{N}}$, this picture is useful for investigating a crossover between
polynomial and stretched exponential scaling. Since the stretched exponential
scaling is the consequence of central limit theorem, weaker disorder (smaller $\operatorname{var} J$) means much larger sizes are needed to observe exponential
complexity.

\section{\label{app:groundstate}Ground state probability}

Observe that creation of all $N$ elementary excitations annihilate
($b_0^{\dag} b_1^{\dag} \cdots b_{N - 1}^{\dag} | \Psi \rangle = 0$) 
all but the ground state $| \Psi \rangle = | 0 \rangle$. Therefore,
the probability to find the system in its ground state at the end of the
annealing algorithm can be expressed as
\begin{equation}
  P_0 = \langle \Psi (T) | b_0 b_0^{\dag} b_1 b_1^{\dag} \cdots b_{N - 1} b_{N
  - 1}^{\dag} | \Psi (T) \rangle .
\end{equation}
Here $b_k$ and $b_k^{\dag}$ are Dirac quasiparticles at $t = T$, which
correspond to kinks (broken bonds between $k - 1$ and $k$; non-zero occupation
number for $k=0$ suggests odd number of kinks)

Since the Hamiltonian is quadratic in fermionic operators we can apply Wick's
theorem to write the success probability as a Pfaffian
\begin{equation}
  P_0 = \operatorname{Pf} \left(\begin{array}{ccccc}
    0 & \langle \Psi | b_0 b_0^{\dag} | \Psi \rangle & \langle \Psi | b_0 b_1
    | \Psi \rangle & \cdots & \langle \Psi | b_0 b_{N - 1}^{\dag} | \Psi
    \rangle\\
    - \langle \Psi | b_0 b_0^{\dag} | \Psi \rangle & 0 & \langle \Psi |
    b_0^{\dag} b_1 | \Psi \rangle & \cdots & \langle \Psi | b_0^{\dag} b_{N -
    1}^{\dag} | \Psi \rangle\\
    - \langle \Psi | b_0 b_1 | \Psi \rangle & - \langle \Psi | b_0^{\dag} b_1
    | \Psi \rangle & 0 & \cdots & \langle \Psi | b_1 b_{N - 1}^{\dag} | \Psi
    \rangle\\
    \vdots & \vdots & \vdots & \ddots & \vdots\\
    - \langle \Psi | b_0 b_{N - 1}^{\dag} | \Psi \rangle & - \langle \Psi |
    b_0^{\dag} b_{N - 1}^{\dag} | \Psi \rangle & - \langle \Psi | b_1 b_{N -
    1}^{\dag} | \Psi \rangle & \cdots & 0
  \end{array}\right) . \label{P0Pf}
\end{equation}
Forming a vector out of Nambu spinors we can express it a linear combination
of Majorana operators
\begin{equation}
  \left(\begin{array}{c}
    b_0\\
    b_0^{\dag}\\
    b_1\\
    b_1^{\dag}\\
    \vdots\\
    b_{N - 1}\\
    b_{N - 1}^{\dag}
  \end{array}\right) = \frac{1}{2} \left(\begin{array}{ccccccc}
    - \ii & 0 & 0 & \cdots & 0 & 0 & - 1\\
    \ii & 0 & 0 & \cdots & 0 & 0 & - 1\\
    0 & 1 & - \ii & \cdots & 0 & 0 & 0\\
    0 & 1 & \ii & \cdots & 0 & 0 & 0\\
    \vdots & \vdots & \vdots & \ddots & \vdots & \vdots &
    \vdots\\
    0 & 0 & 0 & \cdots & 1 & - \ii & 0\\
    0 & 0 & 0 & \cdots & 1 & \ii & 0
  \end{array}\right) \left(\begin{array}{c}
    \chi_0\\
    \chi_1\\
    \chi_2\\
    \chi_3\\
    \vdots\\
    \chi_{2 N - 2}\\
    \chi_{2 N - 1}
  \end{array}\right), \qquad \text{ or, using shorthand notation }
  \boldsymbol{\beta}=\boldsymbol{\mathcal{B}}\boldsymbol{\chi}
  .
\end{equation}
Notice that we assumed antiperiodic boundary conditions in our definition of
$\boldsymbol{\mathcal{B}}$, in accordance with the fact that the number of
fermions is always even.

Recalling the identity $(\operatorname{Pf} X)^2 = \det X$ and exchanging the rows of
the matrix (\ref{P0Pf}) we rewrite in a compact form
\begin{equation}
  P_0 = \sqrt{(- 1)^N \det \langle \Psi |
  (\boldsymbol{\beta}\boldsymbol{\beta}^{\dag} -\boldsymbol{\mathcal{I}}) | \Psi
  \rangle}, \qquad \text{where } \boldsymbol{\mathcal{I}}=
  \left(\begin{array}{ccccccc}
    0 & 0 &  &  &  &  & \\
    0 & 1 &  &  &  &  & \\
    &  & 0 & 0 &  &  & \\
    &  & 0 & 1 &  &  & \\
    &  &  &  & \ddots &  & \\
    &  &  &  &  & 0 & 0\\
    &  &  &  &  & 0 & 1
  \end{array}\right) .
\end{equation}
We need to subtract the matrix $\mathcal{I}$ since annihilation operators
should always appear before the creation operators in the expectation values,
but $b_k$ and $b_k^{\dag}$ do not anticommute. This correction ensures that
the ordering of operators is correct and the argument of the Pfaffian is an
antisymmetric matrix.

It is convenient to use Majorana operators in Heisenberg representation.
Corresponding equations of motion are linear,
\begin{equation}
  \frac{\dd \boldsymbol{\chi}}{\dd t} = 2\mathbf{M} (t) \boldsymbol{\chi}
  \text{ so that } \boldsymbol{\chi} (T) =\mathbf{S}\boldsymbol{\chi} (0),
\end{equation}
where the evolution matrix $\mathbf{S}$ is obtained by integrating a system
of linear differential equations. Then the expectation values are taken with
respect to initial state, i.e. a vacuum,
\begin{equation}
  P_0 = \sqrt{| \det (\boldsymbol{\mathcal{B}}\mathbf{S} \langle 0 |
  \boldsymbol{\chi}\boldsymbol{\chi}^T | 0 \rangle \mathbf{S}^T
  \boldsymbol{\mathcal{B}}^{\dag} -\boldsymbol{\mathcal{I}}) |} .
\end{equation}
Dirac fermions at $t=0$ can be written as a linear combination of Majorana
operators as follows
\begin{equation}
  \left(\begin{array}{c}
     a_0\\
     a_0^{\dag}\\
     a_1\\
     a_1^{\dag}\\
     \vdots\\
     a_{N - 1}\\
     a_{N - 1}^{\dag}
   \end{array}\right) = \frac{1}{2} \left(\begin{array}{ccccccc}
     1 & - \ii & 0 & 0 & \cdots & 0 & 0\\
     1 & \ii & 0 & 0 & \cdots & 0 & 0\\
     0 & 0 & 1 & - \ii & \cdots & 0 & 0\\
     0 & 0 & 1 & \ii & \cdots & 0 & 0\\
     \vdots & \vdots & \vdots & \vdots & \ddots & \vdots & \vdots\\
     0 & 0 & 0 & 0 & \cdots & 1 & - \ii\\
     0 & 0 & 0 & 0 & \cdots & 1 & \ii
   \end{array}\right) \left(\begin{array}{c}
     \chi_0\\
     \chi_1\\
     \chi_2\\
     \chi_3\\
     \vdots\\
     \chi_{2 N - 2}\\
     \chi_{2 N - 1}
   \end{array}\right), \qquad \text{ or, for short, }
   \boldsymbol{\alpha}=\boldsymbol{\mathcal{A}}\boldsymbol{\chi}.
\end{equation}
Using vacuum expectation values
$\langle 0| \boldsymbol{\alpha}\boldsymbol{\alpha}^{\dag} |0 \rangle =
\boldsymbol{1}-\boldsymbol{\mathcal{I}}$ and the identities
$\boldsymbol{\mathcal{A}} \boldsymbol{\mathcal{A}}^\dag =
\boldsymbol{\mathcal{B}} \boldsymbol{\mathcal{B}}^\dag =
\frac{1}{2} \boldsymbol{1}$ we rearrange the terms in the argument of
the determinant function to arrive at the expression used in the main text.

\section{\label{app:numericalproc}Numerical procedure}

To improve the stability of the integrator it is important to ensure that a
solution $\mathbf{S} (t)$ to
\begin{equation}
  \frac{\dd \mathbf{S}}{\dd t} =\mathbf{H}\mathbf{S}, \qquad
  \left( \mathbf{H} \text{ is skew-symmetric} \right)
\end{equation}
is an orthogonal matrix at all times. Runge-Kutta methods of $k$-th order
update the solution as
\begin{equation}
  \mathbf{S}(t + \Delta t)=\mathbf{S}(t) + \Delta \mathbf{S}(t),
\end{equation}
where $\Delta \mathbf{S}(t)$ is chosen so that the approximation error is
$O(\Delta t^{k+1})$. As errors accumulate, orthogonality of
$\mathbf{S}(t)$ is no longer assured.

Methods based on Cayley transform use updates of this form
\begin{equation}
  \mathbf{S}(t+\Delta t) =
  \frac{\boldsymbol{1}+\boldsymbol{\Sigma}/2}{\boldsymbol{1}-\boldsymbol{\Sigma}/2}
  \mathbf{S}(t),
\end{equation}
where $\boldsymbol{1}$ is the identity matrix and $\boldsymbol{\Sigma}$ is
skew-symmetric. It is straightforward to verify that updates of this form
preserve the orthogonality of $\mathbf{S}$. The matrix $\boldsymbol{\Sigma}$
should be chosen so that the approximation error is $O (\Delta t^{k + 1})$.

Methods based on Magnus expansion use updates of the form
\begin{equation}
  \mathbf{S} (t + \Delta t) = \ee^{\boldsymbol{\Omega}} \mathbf{S} (t)
\end{equation}
(where $\boldsymbol{\Omega}$ is skew-symmetric), which similarly maintain
orthogonality. Differential system for $\ln \mathbf{S} (t)$ is a non-linear
one, and leads to an expansion of $\boldsymbol{\Omega}$ in terms of nested
commutators.

The expression is considerably simplified if we observe that $\mathbf{H}$ is
a linear interpolation between constant matrices
$\mathbf{H}= (1-t) \mathbf{A}+ t \mathbf{B}$, where we set $T=1$
without losing generality ($\mathbf{H}$ can be rescaled appropriately).
The time-derivative
$\boldsymbol{\Delta}= \dd \mathbf{H}/ \dd t=\mathbf{B}-\mathbf{A}$
is constant. Further simplification is possible by using a mid-point
of the interval $[t,t+\Delta t]$ so that the evolution is written in a
symmetric form
\begin{equation}
  \mathbf{S} \left( t + \tfrac{\Delta t}{2} \right) =
  \ee^{\boldsymbol{\Omega} (t)} \mathbf{S} \left( t - \tfrac{\Delta t}{2}
  \right).
\end{equation}
The expansion of $\boldsymbol{\Omega}$ can be written in this form
\begin{multline}
  \boldsymbol{\Omega}=\mathbf{H} \Delta t - \frac{1}{12} \mathbf{C} \Delta
  t^3 - \frac{1}{240} [\boldsymbol{\Delta},\mathbf{C}] \Delta t^5 +
  \frac{1}{720} [\mathbf{H},[\mathbf{H},\mathbf{C}]] \Delta t^5 \\
  - \frac{1}{6720} [\boldsymbol{\Delta},[\boldsymbol{\Delta},\mathbf{C}]]
   \Delta t^7 - \frac{1}{30240} [\mathbf{H},[\mathbf{H},[\boldsymbol{\Delta},\mathbf{C}]]] \Delta t^7 + \frac{1}{7560}
   [\boldsymbol{\Delta},[\mathbf{H},[\mathbf{H},\mathbf{C}]]] \Delta t^7 -
   \frac{1}{30240} [\mathbf{H},[\mathbf{H},[\mathbf{H},[\mathbf{H},
   \mathbf{C}]]]] \Delta t^7,
\end{multline}
where $\mathbf{C}= [\mathbf{H}, \boldsymbol{\Delta}] = 
[\mathbf{A}, \mathbf{B}]$. The approximation error for this method is
$O(\Delta t^9)$, i.e. the method is 8th order.

The method based on Magnus expansion is impractical since matrix
exponentiation is expensive. However, maintaining the same level of
precision, we can express Pad\'e approximant of the exponential as a
sequence of Cayley transforms
\begin{equation}
  \ee^ {\boldsymbol{\Omega}} = \prod_{k=1}^4
  \frac{\sigma_k\boldsymbol{1}+\boldsymbol{\Omega}}{\sigma_k\boldsymbol{1}-\boldsymbol{\Omega}}
  + O(\Delta^9),
\end{equation}
where $\sigma_1,\ldots,\sigma_4$ are the roots of
$\sigma^4-20\sigma^3+180\sigma^2-840\sigma+1680=0$.
Lower order methods are obtained by truncating the commutator series
earlier and using lower-order polynomials.

Since $\mathbf{H}$ is a tridiagonal matrix, the bandwidth of
$\boldsymbol{\Omega}$ is $O(p)$ where $p$ is the order of the method.
Using band-diagonal representation the computational cost of a single
step is only $O(p^3 N)$ corresponding to sparse multiplication and
band-diagonal solve.

We also use embarrassing parallelization to speed-up the computation. One approach is to evolve a subset of columns of
$\mathbf{S}$ independently. A complementary approach is to divide the time interval $[0,T]$ can into $n$ subintervals and independently integrate equations of motion using the identity matrix as the initial condition ($\mathbf{S_k}((k-1) T/n)=\boldsymbol{1}$).
The reduction step multiplies the matrices together
$\mathbf{S}(T)=\mathbf{S}_n(T) \mathbf{S}_{n-1}((n-1)T/n)
  \cdots \mathbf{S}_1(T/n)$. The latter cost grows more rapidly
with size, as $O(N^3)$ but does not scale with $T$, so parallelization
is justified for longer evolution times.

\begin{figure}[!ht]
  \includegraphics[width=0.5\linewidth]{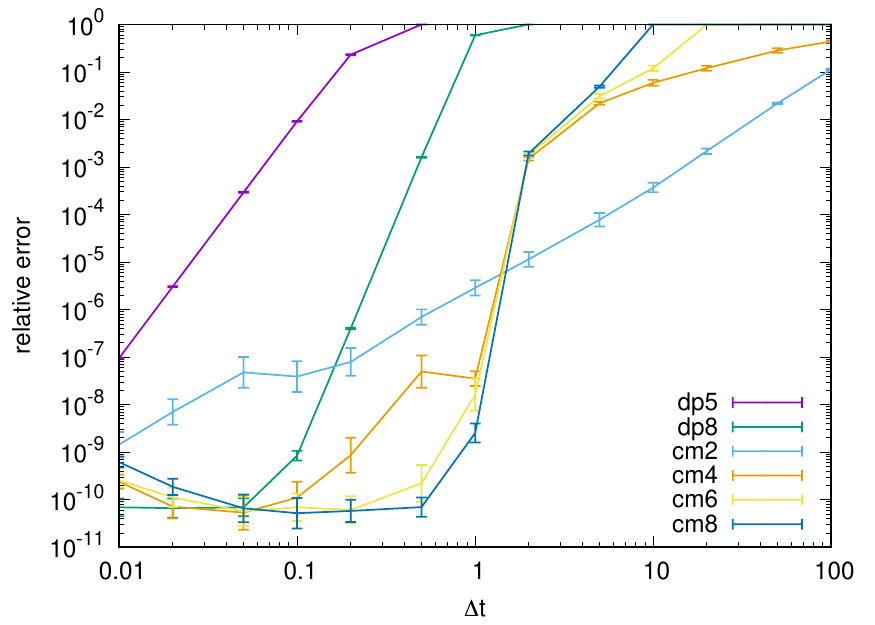}
  \caption{\label{fig:bench}
    Median value (obtained from 10 random instances for $N=64$ and $T=4096$)
    of the relative error vs. step size using Dormand-Prince 5th and 8th order methods (dp5,dp8)
    and Cayley-Magnus method of 2nd, 4th, 6th, and 8th order (cm2,cm4,cm6,cm8).
  }
\end{figure}

Fig.~\ref{fig:bench} compares the relative errors for ground state
probability of random instances with $N=64$ and $T=4096$ 
as a function of step size, obtained using various integrators.
Cayley-Magnus method achieves the best performance.
Python code (version 3.5 or above, using NumPy and SciPy packages)
implementing the integration method is presented on the next page.

\newpage
\subsection*{Python reference implementation}
\begin{verbatim}
import sys
from numpy import *
from scipy import sparse as sp, linalg as la

# Banded matrix multiplication
def bxb(A,B):
    d=(size(A,0)-1)//2
    C=A*B[d,:]
    for s in r_[-d:0]: C[:s,-s:]+=A[-s:,:s]*B[d+s,-s:]
    for s in r_[:d]+1: C[s:,:-s]+=A[:-s,s:]*B[d+s,:-s]
    return C

# Main routine
def solve(G,J,t1,t2,T,p=4,dt=1.0):
    if (J[0]!=0): sys.exit('Use open BCs')
    if (p>4): sys.exit('Use p<=4')

    N=len(J)
    k=r_[:p]; s=roots(r_[1,cumprod(-(p+k+1)*(p-k)/(k+1))])

    t1/=T; t2/=T; dt/=T; G*=T; J*=T
    dt=(t2-t1)/ceil((t2-t1)/dt)

    d=2*p-1
    I=zeros((2*d+1,2*N)); I[d,:]=1
    A=zeros((2*d+1,2*N)); A[d-1,1::2]=2*G; A[d+1,::2]=-2*G
    B=zeros((2*d+1,2*N)); B[d-1,2::2]=2*J[1:]; B[d+1,1:-1:2]=-2*J[1:]
        
    D=(B-A)*dt**2
    C=(bxb(A,B)-bxb(B,A))*dt**3
    
    S=eye(2*N); t=t1+dt/2
    while t<t2:
        H=((1-t)*A+t*B)*dt
        W=H
        if p>=2:
            W+=-(1/12)*C
        if p>=3:
            DC=bxb(D,C)-bxb(C,D)
            HC=bxb(H,C)-bxb(C,H)
            HHC=bxb(H,HC)-bxb(HC,H)
            W+=-(1/240)*DC+(1/720)*HHC
        if p>=4:
            DDC=bxb(D,DC)-bxb(DC,D)
            HDC=bxb(H,DC)-bxb(DC,H)
            HHDC=bxb(H,HDC)-bxb(HDC,H)
            DHHC=bxb(D,HHC)-bxb(HHC,D)
            HHHC=bxb(H,HHC)-bxb(HHC,H)
            HHHHC=bxb(H,HHHC)-bxb(HHHC,H)
            W+=-(1/6720)*DDC-(1/30240)*HHDC+(1/7560)*DHHC-(1/30240)*HHHHC
        for sk in s:
            S=sp.spdiags(sk*I+W,r_[d:-d-1:-1],2*N,2*N)@S
            S=la.solve_banded((d,d),sk*I-W,S)
        S=real(S)
        t+=dt
    return U
\end{verbatim}
\end{document}